\documentstyle[epsfig]{article}

\makeatletter%
\def\nottoobig#1{{\hbox{$\left#1\vcenter to1.111\ht\strutbox{}\right.\n@space$}}}
\makeatother%

\newcommand{\redttnp}[1]{ {  {\rm R}_{{#1}{\scriptsize\mbox{-tt}}}^{p}({\np}) }    }

\newcommand{\up}{{\rm UP}}

\newcommand{\p}{{\rm P}}

\newcommand{\co}{{\rm co}}
\newcommand{\np}{{\rm NP}}

\newcommand{\bh}{{\rm BH}}

\newcommand{\pjk}{  {\rm P}^{ {\rm BH}_j : {\rm BH}_k}}

\newcommand{\pkj}{  {\rm P}^{ {\rm BH}_k : {\rm BH}_j}}
\newcommand{\pcd}{  {\rm P}^{{\cal C} : {\cal D}}}
\newcommand{\pdc}{  {\rm P}^{{\cal D} : {\cal C}}}
\newcommand{\ppcd}{  {\rm PP}^{{\cal C} : {\cal D}}}
\newcommand{\ppdc}{  {\rm PP}^{{\cal D} : {\cal C}}}
\newcommand{\pconectwo}{  {\rm P}^{{\cal C}_1 : {\cal C}_2}}
\newcommand{\pctwocone}{  {\rm P}^{{\cal C}_2 : {\cal C}_1}}

\newcommand{\sigmaj}{{\rm \Sigma}_j^p}
\newcommand{\sigmak}{{\rm \Sigma}_k^p}
\newcommand{\sigmai}{{\rm \Sigma}_i^p}

\newcommand{\sigmal}{{\rm \Sigma}_{\ell}^p}
\newcommand{\sigmam}{{\rm \Sigma}_m^p}
\newcommand{\pim}{{\rm \Pi}_m^p}
\newcommand{\sigmajone}{{\rm \Sigma}_{j+1}^p}

\newcommand{\pij}{{\rm \Pi}_j^p}
\newcommand{\pik}{{\rm \Pi}_k^p}

\newcommand{\deltaj}{{\rm \Delta}_j^p}

\newcommand{\psigjk}{  {\p^{\sigmaj : \sigmak}}}
\newcommand{\psigkj}{  {\p^{\sigmak : \sigmaj}}}
\newcommand{\psiglm}{  {\p^{\sigmal : \sigmam}}}

\newcommand{\pcdtt}{ {\p_{1,1\hbox{-}{\rm tt}}^{( {\cal C} , {\cal D})}}}
\newcommand{\pcdT}{ {\p^{{\cal C} , {\cal D}}}}
\newcommand{\psigmakone}{  {\p^{ {{\rm \Sigma}_k^{\it p}}[1]}}}

\newcommand{\pp}{{\rm PP}}
\newcommand{\bpp}{{\rm BPP}}

\newcommand{\conp}{{\rm coNP}}

\newcommand{\sigmatwo}{{\Sigma_2^p}}
\newcommand{\pitwo}{{\Pi_2^p}}

\newcommand{\ph}{{\rm PH}}

\newtheorem{theorem}{Theorem}[section]
\newtheorem{corollary}[theorem]{Corollary}
\newtheorem{definition}[theorem]{Definition}

\newcommand{\proof}{\noindent {\bf Proof:}\quad}

\newcommand{\qedblob}{\mbox{\rule[-1.5pt]{5pt}{10.5pt}}}
\def\literalqed{{\ \nolinebreak\hfill\mbox{\qedblob\quad}}}

\def\qed{\literalqed}

\newtheorem{example}[theorem]{Example}

\newtheorem{proposition}[theorem]{Proposition}
\newcommand{\singlespacing}{\let\CS=
\@currsize\renewcommand{\baselinestretch}{1}\tiny\CS}
\newcommand{\singlespacingplus}{\let\CS=
\@currsize\renewcommand{\baselinestretch}{1.25}\tiny\CS}
\newcommand{\doublespacing}{\let\CS=
\@currsize\renewcommand{\baselinestretch}{1.75}\tiny\CS}
\newcommand{\draftspacing}{\let\CS=
\@currsize\renewcommand{\baselinestretch}{2.0}\tiny\CS}

\makeatother%

\def\pair#1{{{\langle\!\!~#1~\!\!\rangle}}}

\newcommand{\sigmastar}{\mbox{$\Sigma^\ast$}}

\newcommand{\calc}{{\cal C}}
\newcommand{\cald}{{\cal D}}
\newcommand{\calone}{{\cal C}_1}
\newcommand{\caltwo}{{\cal C}_2}

\newcommand{\condition}{\,\nottoobig{|}\:}

\begin{document}

\sloppy

\title{Query Order and the Polynomial Hierarchy} 

\author{%
Edith Hemaspaandra\thanks{\protect Supported in part 
by grant
NSF-INT-9513368/DAAD-315-PRO-fo-ab.  Work done in part while 
visiting 
Friedrich-Schiller-Universit\"at Jena.
Email: {\tt edith@bamboo.lemoyne.edu}.}
\\Department of Mathematics\\
Le Moyne College\\
Syracuse, NY 13214, USA
\and 
Lane A. Hemaspaandra\thanks{\protect Supported in part 
by grants NSF-CCR-9322513 and 
NSF-INT-9513368/DAAD-315-PRO-fo-ab.  Work done in part while 
visiting 
Friedrich-Schiller-Universit\"at Jena.
Email: {\tt lane@cs.rochester.edu}.}
\\Department of Computer Science\\University of Rochester\\
            Rochester, NY 14627, USA
\and 
Harald Hempel\thanks{\protect Supported in part 
by grant
NSF-INT-9513368/DAAD-315-PRO-fo-ab.
Work done in part while 
visiting 
Le~Moyne College.
Email: {\tt hempel@\protect\linebreak[0]informatik.\protect\linebreak[0]uni-jena.de}.}
\\Institut f\"ur Informatik\\
Friedrich-Schiller-Universit\"at Jena\\
07740 Jena, Germany
}

\date{May 27, 1998}

 \maketitle               

\begin{abstract}    
Hemaspaandra, Hempel, and 
Wechsung~\cite{hem-hem-wec:jtoappear:query-order-bh} initiated the 
field of query order, which studies the ways in which
computational power is affected by 
the order in which information sources are accessed.
The present paper 
studies, for the first time, query 
order as it applies to the levels of the polynomial hierarchy.
$\pcd$ denotes the class of languages computable by a polynomial-time 
machine that is allowed one query to ${\cal C}$ followed by one query to 
${\cal D}$~\cite{hem-hem-wec:jtoappear:query-order-bh}.
We prove that the levels of the polynomial hierarchy are order-oblivious:
$\psigjk=\psigkj$.
Yet, we also show that these ordered query classes form new levels in the 
polynomial hierarchy unless the polynomial hierarchy collapses.
We prove 
that all leaf language classes---and thus
essentially all standard complexity classes---inherit all 
order-obliviousness results that hold
for~P.
\end{abstract}

\section{Introduction}\label{s:intro}
Query order was introduced by Hemaspaandra, Hempel, and 
Wechsung~\cite{hem-hem-wec:jtoappear:query-order-bh} in order to study whether 
the order in which
information sources are accessed has any effect on the class of
problems that can be solved.  In the everyday world, the
order in which we access information is crucial, 
and the work of
Hemaspaandra, Hempel, and Wechsung~\cite{hem-hem-wec:jtoappear:query-order-bh}
shows that this 
real-world intuition holds true in complexity
theory when the
information one is accessing is from the 
boolean hierarchy.
In particular, 
let $\pcd$ denote the class of languages $L$ such that, for some 
$C \in {\cal C}$ and some $D \in {\cal D}$, $L$ is accepted by some \p\ 
transducer $M$ that on any input may make at most one query to $C$ followed 
by at most one query to $D$.
Hemaspaandra, Hempel, and Wechsung
show that, unless the polynomial  
hierarchy collapses, query order always matters 
when $\cal C$ and $\cal D$ are nontrivial 
levels of the boolean 
hierarchy~\cite{cai-gun-har-hem-sew-wag-wec:j:bh1}, except in two cases.
In particular they prove that,
for $1\leq j \leq k$, $\pjk=\pkj$ if $$j=k \mbox{ or } (j
\mbox{ is even and } k=j+1),$$ 
and they prove that unless the polynomial hierarchy 
collapses these are the only cases 
(satisfying $1\leq j \leq k$) for which $\pjk=\pkj$.

The goal of the present paper is to study query order in the 
{\em polynomial\/} hierarchy.
Section \ref{s:ph} shows that, in sharp contrast with the case of the boolean 
hierarchy, {\em query order never matters in the polynomial hierarchy}:
For any $j$ and $k$, $\psigjk=\psigkj$.
We prove this by providing 
for ``$\pcd = \pdc$'' a sufficient condition,
which also has applications in other settings.

Of course, if for $j \leq k$, $\psigjk=\psigmakone$, then our 
$\psigjk=\psigkj$ theorem would be trivial. 
Here, as is standard, $\psigmakone$ denotes the class of 
languages that are computable via polynomial-time 
1-Turing reductions to $\sigmak$~\cite{lad-lyn-sel:j:com}.
In fact, the statement 
$\psigjk=\psigmakone$, for $j < k$, might on casual consideration seem 
plausible, as certainly a $\sigmak$ oracle can simulate the 
$\sigmaj$ query (when $j <k$) of $\psigjk$, can compute the 
answer to it, and then can based on the answer determine the 
$\sigmak$ query of $\psigjk$ and can simulate it.
(Footnote~\ref{f:why-wrong} explains why this 
argument fails to establish
$\psigjk=\psigmakone$.)  
Nonetheless, 
we show that, unless the
polynomial hierarchy collapses, $\p^{\sigmaj:\sigmak} = \p^{\sigmal:\sigmam}$
only if 
$\{j,k\}=\{\ell,m\}$.

In Section~\ref{s:other-base}, we show that {\em all\/} query order 
exchanges that hold for $\pcd$---not just all those we prove but rather 
all that are true---are automatically inherited by 
all leaf language classes, and thus by
essentially all standard complexity classes.
This shows that 
{\em these classes allow at least as many query order exchanges as 
\p\ does}.
We also note that some of these classes---in particular~$\np$---allow
(unless the polynomial hierarchy collapses) more order 
exchanges than~$\p$ does.

\section{Preliminaries}\label{s:prelim}

For standard notions not defined here, we refer
the reader to any computational complexity
textbook, 
e.g.,~\cite{bov-cre:b:complexity,bal-dia-gab:b:sctI-2nd-ed,pap:b:complexity}.

We say a set is trivial if it is $\emptyset$ or $\sigmastar$, and 
otherwise we say it is nontrivial.
A complexity class is any collection of subsets of 
$\sigmastar$.  For each complexity class $\calc$, 
let $\co\calc$ denote $\{ L \condition \overline{L} \in \calc\}$.
The polynomial hierarchy is defined as follows: $\Sigma_0^p = \Pi_0^p =
\Delta_0^p = \Delta_1^p = \p$ and, for 
each $i > 0$, $\sigmai =
\np^{\Sigma_{i-1}^p}$, $\Pi_{i}^p = \co\sigmai$,
and $\Delta_i^p = \p^{\Sigma_{i-1}^p}$.
Let
$A \oplus B$ denote the disjoint union of the sets $A$ and $B$, i.e.,
$A \oplus B = \{x0 \condition x \in A\} \cup \{x1 \condition x \in B\}$,
and let $A \times B$ denote the 
Cartesian product of the sets $A$ and $B$, i.e.,
$A \times B = \{\pair{x,y} \condition x \in A \mbox{ and } y \in B\}$.

In this paper we use oracles to represent databases that are queried. 
This does not mean that this is a ``relativized worlds'' oracle construction 
paper. 
It is not. 
Rather we use relativization in much the same way 
that it is used to build the 
polynomial hierarchy, namely, relativization by full, natural classes. 

We now present the definitions that will allow us to discuss the
ways---order of access, amount of access, etc.---that databases (modeled
as oracles) are accessed.  We use DPTM as a shorthand for
``deterministic polynomial-time (oracle) Turing machine.'' 
Without loss
of generality, we 
assume that such machines are clocked with clocks that are
independent of the oracle.
$M^A(x)$ denotes the computation of DPTM $M$ with oracle $A$ on input $x$.
On occasion, when the 
oracle involved is clear from
context and we are focusing on the action of $M$,
we may write
$M(x)$ and omit the oracle.

\begin{definition} \label{d:new}
Let ${\cal C}$ and  ${\cal D}$ be complexity
classes.
\begin{enumerate}
\item~\cite{hem-hem-wec:jtoappear:query-order-bh} 
Let $M^{A:B}$ denote DPTM $M$ restricted
to making at most one query to oracle $A$
followed  by at most one query to oracle $B$. 
$$\pcd=
\{L \subseteq \Sigma^* \condition (\exists C \in {\cal C})(\exists D \in 
{\cal D})(\exists ~ {\rm DPTM }~M)[L=L(M^{C : D})]\}.$$

\item  \label{p:d:NOWselivanov}
Let $M_{1,1\hbox{-}{\rm tt}}^{(A,B)}$
denote DPTM $M$
restricted to 
making  simultaneously at most one query to oracle $A$ and at most one query
to oracle
$B$.
$${\p_{1,1\hbox{-}{\rm tt}}^{({\cal C},{\cal D})}}=
\{L \subseteq \Sigma^* \condition (\exists C \in {\cal C})(\exists D \in 
{\cal D})(\exists ~ {\rm DPTM}~M)[L=L(M_{1,1\hbox{-}{\rm tt}}^{(C,D)})]\}.$$

\item 
Let $M^{A,B}$ denote DPTM $M$ restricted to making 
at most one query to oracle $A$ and at most one query to
oracle $B$, in arbitrary order. Similarly, let 
$M^{A[1],B[poly]}$ denote DPTM $M$ making 
at most one query to oracle $A$ and polynomially many queries to $B$,
in arbitrary order (it is even legal 
for the query to $A$ to be sandwiched between 
queries to $B$).
$$\p^{{\cal C} , {\cal D}}=
\{L \subseteq \Sigma^* \condition (\exists C \in {\cal C})(\exists D \in 
{\cal D})(\exists ~ {\rm DPTM }~M)[L=L(M^{C , D})]\}.$$
$\p^{{\cal C}[1]  , {\cal D}[poly]}= $ \nopagebreak
$$\{L \subseteq \Sigma^* \condition (\exists C \in {\cal C})(\exists D \in 
{\cal D})
(\exists ~ {\rm DPTM}~M)[L=L(M^{C[1] , 
D[poly]})]\}.$$
\end{enumerate}
\end{definition}

As has been noted by the authors 
elsewhere~\cite{hem-hem-hem:cOutByJournalwJtag:sn-1tt-np-completeness},
part~\ref{p:d:NOWselivanov} 
of Definition~\ref{d:new} is somewhat related to work of
Selivanov~\cite{sel:c:refined-PH}.  
Independently of~\cite{hem-hem-hem:cOutByJournalwJtag:sn-1tt-np-completeness},
Klaus Wagner~\cite{wag:jtoappear:parallel-difference}
has made similar observations in a more general
form (namely, applying to more than two sets and to more 
abstract classes) regarding the relationship between Selivanov's 
classes and parallel-access classes.
For completeness, we repeat 
here, as the present paragraph, 
some text from~\cite{hem-hem-hem:cOutByJournalwJtag:sn-1tt-np-completeness}
that presents the 
basic facts 
known
about the relationship between 
the classes of Selivanov (for the case of 
``$\triangle$''s of two sets;  see 
Wagner~\cite{wag:jtoappear:parallel-difference} 
for the case of more than two sets) and the classes discussed in
this paper.
Selivanov
studied refinements of the
polynomial hierarchy.
Among the classes he considered, those closest 
to the classes we study in this paper are his 
classes
$$ \sigmai \mbox{\boldmath$\bf \triangle$} \sigmaj
=
\{ L \condition 
(\exists A \in \sigmai)
(\exists B \in \sigmaj)
[ L = A \triangle B]\},$$
where $A \triangle B = (A - B) \cup (B - A)$.
Note, however, that his classes seem to be
different from our classes.  
This can be immediately seen from the fact that all our 
classes are closed under complementation, but the
main theorem of Selivanov
(\cite{sel:c:refined-PH}, see also the 
discussion and
strengthening in~\cite{hem-hem-hem:tSPECIAL:translating-downwards})
states that  no 
class of the form 
$\sigmai 
\mbox{\boldmath$\bf \triangle$} \sigmaj$,
with $i>0$ and $j>0$, 
is closed under complementation unless the 
polynomial hierarchy collapses.  Nonetheless,
the class
$\sigmai 
\mbox{\boldmath$\bf \triangle$} \sigmaj$
is not too much weaker than
$\p_{1,1\hbox{\rm{}-}\rm{}tt}^{(\sigmai,\sigmaj)}$,
as it is 
not hard to see (by 
easy manipulations 
if 
$i \neq j$, and from the work of 
Wagner~\cite{wag:j:bounded} and K\"obler,
Sch\"oning, and Wagner~\cite{koe-sch-wag:j:diff}
for the $i=j$ case)
that, for all $i$
and $j$, it holds that $\{ L \condition (\exists L' \in
\sigmai 
\mbox{\boldmath$\bf \triangle$} \sigmaj) [L
\leq_{1\hbox{\rm{}-}\rm{}tt}^p L']\} = 
\p_{1,1\hbox{\rm{}-}\rm{}tt}^{(\sigmai,\sigmaj)}$.
Here, as is standard, $\leq_{1\hbox{\rm{}-}\rm{}tt}^p$
denotes polynomial-time 1-truth-table 
reducibility~\cite{lad-lyn-sel:j:com}.

Let ${\cal C}$ be a complexity class.  In the 
literature,
$\leq_m^p$ denotes many-one polynomial-time
reducibility.  Similarly,
we write 
$A \leq_m^{p,{\cal C}[1]} B$ 
if and only if there is a 
(total) function $f \in {\rm FP}^{{\cal C}[1]}$ such that, 
for all $x$, $x\in A \iff f(x) \in B$.

\section{Query Order in the Polynomial Hierarchy}\label{s:ph}

\subsection{Order Exchange in the Polynomial Hierarchy}
\label{subs:orderex}
We first state and prove a sufficient condition for order
exchange. This condition will apply to a large number of classes.
\begin{theorem}
\label{t:order-exchange-ADDEDdisjointUnionHypothesis} 
If ${\cal C}$ and ${\cal D}$ are classes such that 
${\cal C}$ is closed under 
disjoint union and 
${\cal C}$ is closed downwards
 under $\leq_m^{p,{\cal D}[1]}$,
then $$\pcd = \pdc = \pcdtt.$$
\end{theorem}

Proposition~\ref{p:down} 
notes that for complexity classes that have complete sets, closure
under disjoint union follows from downward closure under many-one
reductions.  For most standard classes $\calc$
this proposition can be used, when applying various theorems of this 
section, to remove
the condition that $\calc$ be closed under disjoint union.

\begin{proposition} \label{p:down}
If $\calc$ has $\leq_m^p$-complete sets and $\calc$ is closed
downwards under $\leq_m^p$-reductions, then $\calc$ is closed under
disjoint union.
\end{proposition}

Before proving Theorem~\ref{t:order-exchange-ADDEDdisjointUnionHypothesis} we 
first prove some results that will be helpful in 
the proof.  Also, Theorem~\ref{t:pcd-in-pdc} 
may apply even in some cases where 
Theorem~\ref{t:order-exchange-ADDEDdisjointUnionHypothesis}'s
hypothesis does not hold.

\begin{definition}
We say ${\cal C}$ ``ands'' $({\cal C},{\cal D})$ if 
$(\forall C \in {\cal C})(\forall D \in {\cal D}) [C \times D 
\in {\cal C}]$.
\end{definition}

\begin{theorem} \label{t:pcd-in-pdc}
If ${\cal C}$ is closed under disjoint union,
${\cal C}$ ``ands'' $({\cal C},{\cal D})$,
and ${\cal C}$ ``ands'' 
$({\cal C},{\rm co{\cal D}})$, then $\pcd \subseteq \pdc$.
\end{theorem}

\proof
Suppose $L \in \pcd$ and let DPTM $M$, $C \in  {\cal C}$, and 
$D \in {\cal D}$ be such that $L=L(M^{C : D})$. Without loss of 
generality, let $M$ always query each of $C$ and $D$ 
exactly once,
regardless of the answer of the  first query (that is,
even given an incorrect answer to the first query, $M$ will
always ask a second query).
We describe a DPTM $N$ and a set $C'$ such that $C' \in {\cal C}$
and $L=L(N^{D:C'})$. Let $C' = \left( (C 
\times \overline{D}) \oplus (C \times D) \right)
\oplus C$, i.e., $C'=$
$$\{\pair{y_1,y_2}00  \condition y_1 \in C \mbox{ and } y_2 \not \in D\} \cup
\{\pair{y_1,y_2}10  \condition y_1 \in C \mbox{ and } y_2 \in D\} \cup
\{y1  \condition y \in C \}.$$
On input $x$, DPTM $N^{D:C'}$ works as follows:
\begin{enumerate}
\item It determines the first and the two potential second queries of 
$M(x)$. Denote the first query by $q_0$ and the two potential second 
queries by $q_y$ and $q_n$, where $q_y$ is the query asked by 
$M(x)$ if the first query was answered ``yes,'' and $q_n$ the query asked
if the first query was answered ``no.''
\item $N$ queries $q_n$ to $D$.
\item  $N$ determines the truth-table of $M(x)$ 
with respect to $q_0$ and $q_y$, with query $q_n$ answered
correctly.  That is, let
$(X_1,X_2,X_3), X_i \in \{{\rm A,R}\}$,
where A stands for accept and R for reject, be such that
\begin{enumerate}
\item $X_1$ is the outcome of $M(x)$ if both $q_0$ and $q_y$ are answered
 ``yes'' (recall that if $q_0$ is answered ``yes'' then
$M(x)$ asks $q_y$ as its second query),
\item $X_2$ is the outcome of $M(x)$ if $q_0$ is answered 
``yes'' and $q_y$ is answered  ``no,'' and 
\item $X_3$ is the outcome of $M(x)$ if $q_0$ is answered 
``no'' and $q_n$ is answered correctly.
\end{enumerate}
\item There are eight
different cases for $(X_1,X_2,X_3)$. We have to show that
each case can be handled in polynomial time with one query to $C'$.
We will henceforward assume
that there are more Rs than As in $(X_1,X_2,X_3)$. (The remaining cases follow
by complementation.) Depending on the determined truth-table $(X_1,X_2,X_3)$, 
$N$ does the following: 
\begin{enumerate}
\item $(X_1,X_2,X_3) = ({\rm R, R, R})$. In this case, $N$ will of course
reject.
\item $(X_1,X_2,X_3) = ({\rm A, R, R})$. Then $M$ accepts if and only if
$q_0 \in C$ and $q_y \in D$. This is the case if and only if 
$\pair{q_0,q_y}10 \in C'$. So $N$ queries $\pair{q_0,q_y}10$ to $C'$ and accepts
if and only if the answer is ``yes.''
\item $(X_1,X_2,X_3) = ({\rm R, A, R})$. Then $M$ accepts if and only if
$q_0 \in C$ and $q_y \not \in D$. This is the case if and only if 
$\pair{q_0,q_y}00 \in C'$. So $N$ queries $\pair{q_0,q_y}00$ to $C'$ and accepts
if and only if the answer is ``yes.''
\item $(X_1,X_2,X_3) = ({\rm R, R, A})$. Then $M$ accepts if and only if
$q_0 \not \in C$. This is the case if and only if $q_01 \not \in C'$. So $N$
queries
$q_01$ to $C'$ and accepts if and only if the answer is ``no.''
\end{enumerate}
\end{enumerate}

It is clear from the construction that $L(M^{C : D})=L(N^{D:C'})$ 
and thus $L \in \pdc$, since $C' \in {\cal C}$ by the 
closure properties in the theorem's hypothesis.~\qed

\begin{corollary} \label{c:pcd-in-pdc-ADDEDDisjointUnionHypothesis}
If ${\cal C}$ and ${\cal D}$ are classes such that 
${\cal C}$ is closed under disjoint union and
${\cal C}$ is closed downwards under $\leq_m^{p,{\cal D}[1]}$,
then $$\pcd \subseteq \pdc.$$
\end{corollary}

\proof 
If ${\cal C}$ contains only trivial sets, i.e., ${\cal C}
\subseteq \{\emptyset, \sigmastar\}$, then $\pcd = {\rm P}^{\cal D} = \pdc$
and we are done.  So from now on we assume
that ${\cal C}$ contains a nontrivial set. We will show that
in this case we can apply Theorem~\ref{t:pcd-in-pdc}, i.e., we will show
that  ${\cal C}$, which is closed under disjoint union, 
has also the properties that 
${\cal C}$ ``ands'' $({\cal C},{\cal D})$
and ${\cal C}$ ``ands'' $({\cal C},{\rm co{\cal D}})$.

Let $C \in {\cal C}$ 
and $D \in {\cal D}$. We need to show that $C \times D \in {\cal
C}$ and $C \times \overline{D} \in {\cal C}$. 
If $C \neq \sigmastar$, then $C \times D \leq_m^{p,{\cal D}[1]}
C$ by $f(\pair{x,y}) = x$ if $y \in D$ and some fixed element not in $C$ if $y
\not \in D$. Since ${\cal C}$ is closed under $\leq_m^{p,{\cal
D}[1]}$, it follows that $C \times D \in {\cal C}$. 
Similarly, if $C \neq \sigmastar$, then $C \times \overline{D}
\in {\cal C}$.

If $\sigmastar \in {\cal C}$, 
it remains to show that
$\sigmastar \times D$ and $\sigmastar \times \overline{D} \in {\cal C}$.
Let $C \in {\cal C}$ be a nontrivial set
(recall that we earlier eliminated
the case in which $\cal C$ lacks 
nontrivial sets), and let $c \in C$ and $\widehat{c} \not \in
C$. Then $\sigmastar \times D \leq_m^{p,{\cal D}[1]}
C$ by $f(\pair{x,y}) = c$ if $y \in D$ and $\widehat{c}$ if $y \not \in D$,
and $\sigmastar \times \overline{D} \leq_m^{p,{\cal D}[1]}
C$ by $f(\pair{x,y}) = c$ if $y \not \in D$ and $\widehat{c}$ 
if $y  \in D$.~\qed

\noindent{\em Proof of 
Theorem~\ref{t:order-exchange-ADDEDdisjointUnionHypothesis}.}~~Let 
${\cal C}$ and ${\cal D}$ be classes such that 
${\cal C}$ is closed under disjoint union
and $\calc$ is closed downwards under $\leq_m^{p,{\cal D}[1]}$.
We have to show that $\pcd = \pdc = \pcdtt.$
The 
containment
$\pcd \subseteq \pdc$ follows from 
Corollary~\ref{c:pcd-in-pdc-ADDEDDisjointUnionHypothesis}, and
$\pcdtt \subseteq \pcd$ is immediate from the definitions.

It remains to show that
$\pdc \subseteq \pcdtt$.
Suppose $L \in \pdc$ and let DPTM $M$, $C \in  {\cal C}$, and 
$D \in {\cal D}$ be such that $L=L(M^{D : C})$. Without loss of 
generality, let $M$ always query both $D$ and $C$.
We now describe a DPTM $N$ and a set $C'$ such that $C' \in {\cal C}$
and $L=L(N_{1,1\hbox{-}{\rm tt}}^{(C',D)})$. Define 
\[C'=\{x \condition \mbox{the second query asked by 
$M^{D : C}(x)$ is in } C\}.\]
Since ${\cal C}$ is closed downwards under $\leq_m^{p,{\cal D}[1]}$,
we clearly have $C' \in {\cal C}$.

Let $N_{1,1\hbox{-}{\rm tt}}^{(C',D)}$ on input $x$ work as follows:
$N_{1,1\hbox{-}{\rm tt}}^{(C',D)}(x)$ simulates  $M^{D:C}(x)$ 
until $M^{D:C}(x)$ asks its first query, call it $q$. 
Then  $N_{1,1\hbox{-}{\rm tt}}^{(C',D)}(x)$ queries 
``$x \in C'?$''\ and
``$q \in D$?''~~$N$ at this 
point has enough information to 
simulate the final action of $M$.  We make this completely
rigorous and formal as follows.
Let $S_{C'}$ be $\sigmastar$ if $x \in C'$ and 
let $S_{C'}$ be $\emptyset$ if $x \not\in C'$.
Let $S_{D}$ be $\sigmastar$ if $q \in D$ and 
let $S_{D}$ be $\emptyset$ if $q \not\in D$.
$N_{1,1\hbox{-}{\rm tt}}^{(C',D)}(x)$ accepts if and only if 
$M^{S_D:S_{C'}}(x)$ accepts (which $N(x)$ can easily 
determine given the answers to $N(x)$'s two queries).
It is clear from the construction that 
$L(M^{D:C})=L(N_{1,1\hbox{-}{\rm tt}}^{(C',D)})$, and thus 
$L \in \pcdtt$.~\qed

In addition to leading to the ``polynomial hierarchy
is order-oblivious'' results that this section will
obtain, and leading to Section~\ref{s:other-base}'s
applications to probabilistic and unambiguous 
classes, Theorem~\ref{t:order-exchange-ADDEDdisjointUnionHypothesis} 
has also 
played an important role in distinguishing 
robust Turing and many-one 
completeness~\cite{hem-hem-hem:cOutByJournalwJtag:sn-1tt-np-completeness}.

The next theorem shows that if ${\cal C}$ and
${\cal D}$ are closed under disjoint union and 
are order-oblivious with respect to $\p$
transducers, then
ordered access equals arbitrary-order access.
Note that Theorem~\ref{t:order-T}'s hypothesis 
requires that {\em both\/} classes be closed under
disjoint union, in contrast to the hypothesis of
Theorem~\ref{t:pcd-in-pdc}.

\begin{theorem}\label{t:order-T}
If $\calc$ and $\cald$ are complexity classes that are both closed under 
disjoint union, then
$$\pcd=\pdc \Rightarrow \pcd = \pcdT.$$
\end{theorem}

\proof
Suppose that $\pcd=\pdc$. Since 
$\pcd \subseteq \pcdT$,  
we have only to show that $\pcdT \subseteq \pcd$.
Let $L \in \pcdT$, and 
let DPTM $M$, $C \in \calc$, and 
$D \in \cald$ be such that  $L = L(M^{C,D})$. Without loss of
generality, we assume
that $M^{C,D}$ always queries each oracle exactly once.
Define 
$$L_1=\{x \in L \condition M^{C,D}(x) \mbox{ first queries }C\}.$$
$$L_2=\{x \in L \condition M^{C,D}(x) \mbox{ first queries }D\}.$$
Let $N$ be a DPTM such that $L_1=L(N^{C:D})$.
Since clearly $L_2 \in \pdc$, by our hypothesis 
there exists a DPTM $T$, and sets $C' \in \calc$ and
$D' \in \cald$, such that $L_2=L(T^{C':D'})$.
Let $\widehat{C} = C \oplus C'$ and $\widehat{D}=D\oplus D'$.
We describe a DPTM $S$ such that $L=L(S^{\widehat{C}:\widehat{D}})$.

$S$  on input $x$ will work as follows:
$S(x)$ simulates 
(appropriately tagging a 0 or a 1 onto the 
end of queries to address the 
appropriate part of the disjoint union)
$M^{C,D}(x)$ until $M^{C,D}(x)$ makes its first 
query. 
Then $S(x)$ simulates $N^{C:D}(x)$ or  
$T^{C':D'}(x)$, depending on whether the first 
query of $M^{C,D}(x)$ was asked to $C$ or $D$, respectively.
Note that clearly $L=L(S^{\widehat{C}:\widehat{D}})$, and thus  
$L \in \pcd$.~\qed

From Theorem~\ref{t:order-exchange-ADDEDdisjointUnionHypothesis} 
and Theorem~\ref{t:order-T}
we have the following.
\begin{corollary}
\label{c:order} If ${\cal C}$ and ${\cal D}$ are classes such that 
${\cal C}$ is closed downwards under $\leq_m^{p,{\cal D}[1]}$ and 
${\cal C}$ and ${\cal D}$ are
closed under disjoint union,
then $$\pcdtt = \pcd = \pdc = \pcdT.$$
\end{corollary}

Corollary~\ref{c:order}
implies that query  order does not matter in the polynomial
hierarchy.

\begin{corollary}\label{c:sig}
\begin{enumerate}
\item \label{sig:parta}For all $j,k \geq 0$,  
$\psigjk=\psigkj.$
\item \label{sig:partb}For all $j,k \geq 0$ such that $j \not= k$,
$\p_{1,1\hbox{-}{\rm tt}}^{(\sigmaj,\sigmak)} = \p^{\sigmaj: \sigmak} =
\p^{\sigmaj, \sigmak}$.
\end{enumerate}
\end{corollary}

\proof
Note that for $j=k$ the claim of part~\ref{sig:parta} is trivial. 
Assume $j<k$ (the $j>k$ case is similar).
It is immediately clear that $\sigmak$ is closed downwards under 
$\leq_m^{p,\sigmaj[1]}$ and it is well-known
that
$\sigmaj$ and $\sigmak$ are both closed under disjoint union.
So we can apply Corollary~\ref{c:order}.  Thus, both
parts of the theorem are established.~\qed

Note that in part~\ref{sig:partb} of Corollary~\ref{c:sig} we need 
$j \not= k$, since otherwise we would have included the claim that two 
truth-table queries to $\sigmak$ have as much computational power as two 
Turing queries. 
However, that would imply that the boolean hierarchy over $\sigmak$
collapses to the 2-truth-table closure of $\sigmak$, which in turn
would 
imply
that the polynomial hierarchy 
collapses.
The last implication refers to the well-known fact
that if the boolean hierarchy collapses then the 
polynomial hierarchy collapses;  this fact
was first proven by 
Kadin~\cite{kad:joutdatedbychangkadin:bh}, and the strongest
known collapse of the polynomial hierarchy from a given collapse
of the boolean hierarchy is the one 
recently obtained by
Hemaspaandra et 
al.~\cite{hem-hem-hem:t:easy-hard-survey}
and, independently, by
Reith and Wagner~\cite{rei-wag:tOutByCocoon98:boolean-lowness}.

We also have the following.

\begin{corollary}\label{c:long}
\begin{enumerate}
\item \label{parta} For all $k \geq 0$ and $j > 0$,
$$\p_{1,1\hbox{-}{\rm tt}}^{(\deltaj , \sigmak)} = \p^{\deltaj : \sigmak}
 = \p^{\sigmak : \deltaj} = \p^{\sigmak , \deltaj} = \left\{
\begin{array}{l}
\protect{\deltaj} \mbox{~~~~~~~~~~~~~~~~~if $j>k$}\\
\protect { \p^{\sigmak[1],\Sigma_{j-1}^p[poly]}}
\mbox{~~if $j\leq
k$.}
\end{array}
\right.$$
\item \label{partb} For all $j,k \geq 0$, 
$\p_{1,1\hbox{-}{\rm tt}}^{(\deltaj , \sigmak \cap \pik)}
 = $ \nopagebreak
$$\p^{\deltaj : \sigmak \cap \pik}=\p^{\sigmak \cap \pik : \deltaj} =
\p^{\sigmak \cap \pik , \deltaj} =\left\{
\begin{array}{l}
\protect{\deltaj} \mbox{~~~~~~~if $j>k$}\\
\protect{\sigmak \cap \pik} \mbox{~if 
$j \leq k$.}
\end{array}
\right.$$
\item \label{partc} For all $j,k \geq 0$, 
$\p_{1,1\hbox{-}{\rm tt}}^{(\sigmaj \cap \pij , \sigmak \cap \pik)} =$ 
\nopagebreak
$$\p^{\sigmaj \cap \pij : \sigmak \cap \pik} = 
\p^{\sigmak \cap \pik : \sigmaj \cap \pij} = 
\p^{\sigmak \cap \pik , \sigmaj \cap \pij} = 
\Sigma_{\max(j,k)}^p \cap \Pi_{\max(j,k)}^p.$$
\end{enumerate}
\end{corollary}

\proof
We first prove part~\ref{parta}.
If $j > k$, then a $\deltaj$ machine can simulate $\p^{\deltaj,
\sigmak}$, and it is unconditionally immediate that   
$\p_{1,1\hbox{-}{\rm tt}}^{(\deltaj , \sigmak)}$ 
contains $\deltaj$.
If $0 < j \leq k$, then $\sigmak$ is closed under $\leq_m^{p,\deltaj[1]}$
and thus, by Corollary~\ref{c:order},
$\p_{1,1\hbox{-}{\rm tt}}^{(\deltaj , \sigmak)} = \p^{\deltaj : \sigmak}
 = \p^{\sigmak : \deltaj} = \p^{\deltaj, \sigmak}$. Since $\deltaj =
\p^{\Sigma_{j-1}^p}$, 
$\p^{\sigmak : \deltaj} \subseteq 
\p^{\sigmak[1],\Sigma_{j-1}^p[poly]}$.

It remains to show that
$\p^{\sigmak[1],\Sigma_{j-1}^p[poly]}
\subseteq 
\p^{\sigmak : \deltaj}$.
Suppose $L \in {\p^{\sigmak[1],\Sigma_{j-1}^p[poly]}}$ and let 
DPTM $M$, $A \in \sigmak$, and $B \in \Sigma_{j-1}^p$ be such that 
$L=L(M^{A[1],B[poly]})$.
Without loss of generality, let $M$ ask all its queries to $B$ before asking 
anything to $A$. (If $M$ does not have the desired property, replace
it with a machine that,
before asking anything
to $A$, asks to $B$ the queries $M$ would ask to $B$ if the $A$
query were answered ``yes'' and also 
asks to $B$ the queries $M$ would ask to $B$ if the $A$
query were answered ``no'' and then queries $A$ and uses
the appropriate set of already obtained answers to complete
the simulation of the original $M$.) 
We will denote this with the notation
$L=L(M^{B[poly]:A[1]})$. Also, without loss of generality assume that
$M^{B[poly]:A[1]}$ on input $x$ asks exactly one query $a_x$ to $A$.

Now let us describe a DPTM $N$ and sets $A'$ and $C$ such that  
$A' \in \sigmak$, $C \in \deltaj$, and $L(N^{A':C})=L$.
\begin{itemize}
\item[] $A'=\{x \in \Sigma^* \condition a_x \in A\}$,
\item[] $C=\{x \condition M^{B[poly]: \emptyset[1]}(x) $
accepts$\} \oplus
\{x \condition M^{B[poly]:{\Sigma^{\ast}}[1]}(x) $ accepts
$\}$.
\end{itemize}
Note that the use of $\emptyset$ and $\sigmastar$ in the 
definition of $C$ is just a way to study the effect, respectively,
of ``no'' and ``yes'' oracle answers.
Clearly we have $A' \in \sigmak$ and $C \in \deltaj$.
On input $x$, $N^{A':C}$ will work as follows:
$N^{A':C}(x)$ first queries ``$x \in A'$.'' If the answer 
to ``$x\in A'$'' 
is 
``yes,'' then
$N$  accepts if and only if $x1 \in C$ and if
the answer to ``$x\in A'$'' 
is ``no,'' then
$N$ accepts if and only if $x0 \in C$.
It is immediate that
$L(N^{A':C})=L(M^{B[poly]:A[1]})$ 
and thus $L \in \p^{\sigmak : \deltaj}$.
This completes  the proof of part~\ref{parta} of the corollary.

We now turn to proving part~\ref{partb}.
First note that both $\deltaj$ and $\sigmak \cap
\pik$ are trivially contained in
$\p_{1,1\hbox{-}{\rm tt}}^{(\deltaj , \sigmak \cap \pik)}$.
The $j > k$ case now follows from part~\ref{parta}, since
$\p^{\sigmak \cap \pik , \deltaj} \subseteq \p^{\sigmak, \deltaj}
= \deltaj$.
If $j \leq k$, then a $\sigmak$  machine can simulate $\p^{\deltaj,
\sigmak \cap \pik}$. 
(This simulation is an easy variation of the standard 
simulation showing that $\p^{\sigmak \cap \pik} = \sigmak \cap \pik$,
which itself is a straightforward 
generalization of the early 
work~\cite{sel:j:structure-np,sel:j:pselective-tally}
noting
$\p^{\np \cap \conp} = \np \cap 
\conp$.)~~Since 
$\p^{\deltaj,\sigmak \cap \pik}$ is closed under
complement, it follows that $\p^{\deltaj,\sigmak \cap \pik} \subseteq
\sigmak \cap \pik$.

We now prove part~\ref{partc}.
As in part~\ref{partb}, 
$\sigmaj \cap \pij$ and $\sigmak \cap \pik$ are trivially
contained in 
$\p_{1,1\hbox{-}{\rm tt}}^{(\sigmaj \cap \pij , \sigmak \cap \pik)}$.
Also, a $\Sigma_{\max(j,k)}^p$ machine can (even if $j=k$)
simulate
$\p^{\sigmaj \cap \pij , \sigmak \cap \pik}$ so, by complementation,
we have
$\p^{\sigmaj \cap \pij , 
\sigmak \cap \pik} \subseteq \Sigma_{\max(j,k)}^p \cap
\Pi_{\max(j,k)}^p$.~\qed

\subsection{Query Order Classes Differ from Standard Polynomial Hierarchy 
Levels and from Each Other}\label{subs:differ}

\begin{figure}[tbp]
\begin{center}
\begin{picture}(0,0)%
\epsfig{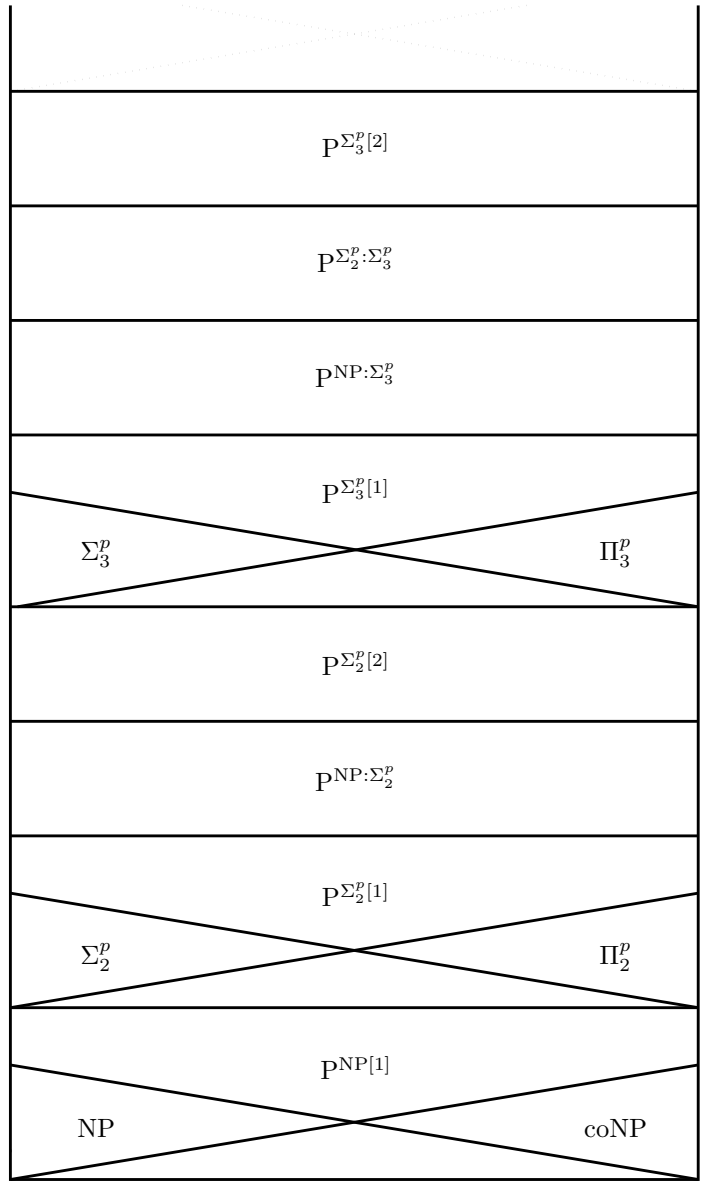}%
\end{picture}%
\setlength{\unitlength}{0.00050000in}%
\begingroup\makeatletter\ifx\SetFigFont\undefined%
\gdef\SetFigFont#1#2#3#4#5{%
  \reset@font\fontsize{#1}{#2pt}%
  \fontfamily{#3}\fontseries{#4}\fontshape{#5}%
  \selectfont}%
\fi\endgroup%
\begin{picture}(7266,12366)(2368,-11794)
\put(3301,-5311){\makebox(0,0)[b]{\smash{\SetFigFont{10}{12.0}{\rmdefault}{\mddefault}{\updefault}$\Sigma_3^p$}}}
\put(6001,-10711){\makebox(0,0)[b]{\smash{\SetFigFont{10}{12.0}{\rmdefault}{\mddefault}{\updefault}$\p^{\np [1]}$}}}
\put(6001,-8911){\makebox(0,0)[b]{\smash{\SetFigFont{10}{12.0}{\rmdefault}{\mddefault}{\updefault}$\p^{\sigmatwo [1]}$}}}
\put(6001,-7711){\makebox(0,0)[b]{\smash{\SetFigFont{10}{12.0}{\rmdefault}{\mddefault}{\updefault}$\p^{\np : \sigmatwo}$}}}
\put(6001,-6511){\makebox(0,0)[b]{\smash{\SetFigFont{10}{12.0}{\rmdefault}{\mddefault}{\updefault}$\p^{\sigmatwo [2]}$}}}
\put(6001,-4711){\makebox(0,0)[b]{\smash{\SetFigFont{10}{12.0}{\rmdefault}{\mddefault}{\updefault}$\p^{\Sigma_3^p [1]}$}}}
\put(6001,-3511){\makebox(0,0)[b]{\smash{\SetFigFont{10}{12.0}{\rmdefault}{\mddefault}{\updefault}$\p^{\np : \Sigma_3^p}$}}}
\put(6001,-2311){\makebox(0,0)[b]{\smash{\SetFigFont{10}{12.0}{\rmdefault}{\mddefault}{\updefault}$\p^{\sigmatwo : \Sigma_3^p}$}}}
\put(6001,-1111){\makebox(0,0)[b]{\smash{\SetFigFont{10}{12.0}{\rmdefault}{\mddefault}{\updefault}$\p^{\Sigma_3^p [2]}$}}}
\put(8701,-11311){\makebox(0,0)[b]{\smash{\SetFigFont{10}{12.0}{\rmdefault}{\mddefault}{\updefault}$\conp$}}}
\put(3301,-11311){\makebox(0,0)[b]{\smash{\SetFigFont{10}{12.0}{\rmdefault}{\mddefault}{\updefault}$\np$}}}
\put(8701,-9511){\makebox(0,0)[b]{\smash{\SetFigFont{10}{12.0}{\rmdefault}{\mddefault}{\updefault}$\pitwo$}}}
\put(3301,-9511){\makebox(0,0)[b]{\smash{\SetFigFont{10}{12.0}{\rmdefault}{\mddefault}{\updefault}$\sigmatwo$}}}
\put(8701,-5311){\makebox(0,0)[b]{\smash{\SetFigFont{10}{12.0}{\rmdefault}{\mddefault}{\updefault}$\Pi_3^p$}}}
\end{picture}
\caption{\protect\label{f:inc}All the classes shown are distinct, unless the polynomial hierarchy 
collapses (see Theorem~\protect\ref{t:differ}).}
\end{center}
\end{figure}

In Section~\ref{s:intro} we mentioned 
that though $\psigjk=\p^{\sigmak[1]}$, 
$0 < j<k$, might seem a tempting 
claim,\footnote{\protect\label{f:why-wrong}{}The reason
the tempting proof implicitly 
sketched in the introduction is not valid 
is that, though $\sigmak$ indeed can simulate $\sigmaj$, $j<k$, 
$\sigmak$ can 
neither pass an extra bit of information back to the ``base'' \p\ machine 
nor---in the crucial case in which the base \p\ machine uses the answer to the 
$\sigmaj$ query to decide whether to treat the $\sigmak$ answer via the 
strictly positive truth-table or the strictly negative truth-table---can it 
complement itself (as that seemingly requires $\sigmak=\pik$). That is, the 
tempting claim fails due to a 1-bit information-passing bottleneck.}
the claim is false unless the polynomial hierarchy collapses.
In fact, we will prove something much stronger, namely that, unless the
polynomial hierarchy collapses, $\p^{\sigmaj:\sigmak} = \p^{\sigmal:\sigmam}$
if and only if $\{j,k\}=\{\ell,m\}$.
The ``if'' 
direction is trivial.  Theorem~\ref{t:differ} establishes 
the ``only if'' direction.

\begin{theorem}
\label{t:differ}
Let $j,k,\ell,m \geq 0$.
If $\psigjk = \psiglm$ then
$\{j,k\}=\{\ell,m\}$
or the polynomial hierarchy collapses.
\end{theorem}

This theorem will follow from a result
of this paper combined with the results and 
techniques of~\cite{hem-hem-hem:tSPECIAL:translating-downwards}.
The following proposition
is a strong and counterintuitive
downward translation result that has recently 
been established.

\begin{proposition} {} \label{p:hhh-NOW-USES-NEW-STRONGER-PAPER-AND-RESULT}
(Special case of \cite[Theorem~2.3]{hem-hem-hem:tSPECIAL:translating-downwards}) \ 
Let $0 < j$ and $j < k$.
If $ \deltaj 
\mbox{\boldmath$\bf \triangle$}
\sigmak = 
\sigmaj
\mbox{\boldmath$\bf \triangle$}
\sigmak$,
then $\sigmak = \pik = \ph.$
\end{proposition}

For all $j$ and $k$, it holds that 
$\psigjk \subseteq \Delta_{j+1}^p 
\mbox{\boldmath$\bf \triangle$}
\sigmak$.  Why?
For $j \geq k$ it is immediate as in that case
$\psigjk \subseteq \p^{\sigmaj [2]}$.
For $j < k$ it follows 
essentially by the technique of 
the {\em proof\/} of
\cite[Lemma~2.3]{hem-hem-hem:jtoappear:downward-translation}.
So, for all $j$ and $k$ we have
$\psigjk \subseteq 
\Delta_{j+1}^p 
\mbox{\boldmath$\bf \triangle$}
\sigmak \subseteq
\sigmajone
\mbox{\boldmath$\bf \triangle$}
\sigmak \subseteq \p^{\sigmajone : \sigmak}$, and thus 
we have the following corollary.

\begin{corollary}\label{c:new-dshandle}
Let $0\leq j$ and $j < k-1$.
If $\psigjk =\p^{\sigmajone:\sigmak}$,
then $\sigmak = \pik = \ph$.
\end{corollary}

We note the strength of this collapse.  The conclusion
obtains a collapse of the hierarchy to a level that is generally
thought to be lower (a priori) than the level of either of the classes whose 
equality was assumed in the hypothesis.  That is, this is an actual
{\em downward\/} translation of equality, in contrast with the 
far more common behavior of upward translation of 
equality (see, 
e.g.,~\cite{wag:t:n-o-q-87version,wag:t:n-o-q-89version,rao-rot-wat:jCheckIfThereIsACorrigedumIThinkThereMayBe:upward}, 
for examples and discussion).

We now can prove Theorem~\ref{t:differ}.

\smallskip

\noindent {\em Proof of 
Theorem~\ref{t:differ}.}~~Suppose that 
$\psigjk = \psiglm$ and that 
(without loss of generality in light of 
Corollary~\ref{c:sig}) $j \leq k$ and  $\ell  \leq m$.
Suppose that either $k < m$ or $(k = m$ and $j < \ell)$.
First note that if $k < m$, then $\psigjk \subseteq \sigmam
\subseteq \psiglm$.  Since $\psigjk = \psiglm$, it follows
immediately that $\sigmam = \pim = \ph$.~~So suppose
that $k=m$ and $j < \ell$.  Then $\psigjk = 
\p^{\sigmajone : \sigmak}$.  If $j < k-1$, then $\sigmak = 
\pik = \ph$ by Corollary~\ref{c:new-dshandle}.  Finally, 
suppose that $j = k-1$.   Then
the class of sets that 2-truth-table reduce to $\sigmak$
sets
equals the 
class of sets that 2-Turing reduce to $\sigmak$
sets, which itself is 
well-known~(\cite{kad:joutdatedbychangkadin:bh},
see 
also
\cite{hem-hem-hem:t:easy-hard-survey,rei-wag:tOutByCocoon98:boolean-lowness})
to imply that~$\ph$ collapses.~\qed

So it is clear from Theorem~\ref{t:differ} that query 
order classes do not equal 
standard ``bounded query'' classes but rather form new intermediate levels in 
the polynomial hierarchy, unless the polynomial hierarchy collapses
(see Figure~\ref{f:inc}).

We conclude this section by mentioning some very recent 
related work that was inspired by the present 
paper.    In this paper, our basic model is of ordered 
access to two sets.  
Wagner~\cite{wag:jtoappear:parallel-difference} 
and 
Beigel and Chang~\cite{bei-cha:c:commutative-queries} 
build
on the work of the current paper by 
studying machines that have ordered access to truth-table groups
of queries and they show that
in that case too
order does not matter.  We consider this work to be 
important and interesting, and to broaden the range of 
models to which the questions of this paper can
be applied.  We also mention that the work is not 
strictly stronger than our work as Beigel and Chang discuss
only sets from the polynomial hierarchy and Wagner has 
somewhat different hypotheses than we do on the classes involved, 
especially regarding our intermediate results that separate out 
exactly what hypotheses imply what conclusions.  Also,
in contrast to the key hierarchy collapse result of the present
paper, which guarantees and proves a downward translation of equality,
the analogous hierarchy collapses of those papers obtain from weaker
hypotheses weaker collapses (namely, the collapse results of those
papers related to query-order-based language classes merely assert
that the hierarchy collapses, and they rely either on the
upward-equality-translation work of Selivanov or make no
specific collapse-level claim at all).  Finally, 
as we will discuss later in more detail,
the work of Section~\ref{s:other-base} applies 
fully to the cases discussed in these papers.  
A survey paper by Hemaspaandra et 
al.~\cite{hem-hem-hem:j:query-order-survey}
provides a detailed overview of query order.

\section{Base Classes Other Than P} \label{s:other-base}

We show that a wide variety of classes inherit 
all order exchanges that hold for $\p$.  For example, if $\pcd \subseteq 
\pdc$ then $\ppcd \subseteq \ppdc$.
Thus order exchanges proven for \p---such 
as those of Section~\ref{subs:orderex} 
of this paper and those of Hemaspaandra, Hempel,
and Wechsung~\cite{hem-hem-wec:jtoappear:query-order-bh}---can immediately
be applied to many other classes.

We prove our result in a very general form, and then state 
some corollaries and 
applications to make the meaning of the theorem more concrete.
For classes ${\cal D}_1$ and ${\cal D}_2$ for which
relativization has been defined, we say
that 
${\cal D}_1$ is robustly contained in ${\cal D}_2$ if, for each $A$,
${\cal D}_1^A\subseteq {\cal D}_2^A$.
${\cal D}^{{\cal C}[1]}$ will mean that each path of the base
machine makes at most one call to ${\cal C}$.
${\cal D}^{{\cal C}_1 : {\cal C}_2 }$ will mean that each path of the base
machine makes at most one call to ${\cal C}_1$
followed by at most one call to ${\cal C}_2$.

\begin{definition}
Let $\cald$ be a complexity class for which relativization is
defined.
We say that $\cald$ is {\em sane\/} if
$$(\forall \calc_1 , \calc_2)[\cald ^ {\calc_1 : \calc_2 }
= \cald ^ {\left({\rm P}^{\calc_1 : \calc_2 }\right) [1]}].$$
\end{definition}

The important point to note is that essentially all standard
complexity classes within the realm of potentially feasible
computation (classes from~P to~PSPACE) are sane.  In particular, 
bringing work of Bovet, Crescenzi, and Silvestri
into notational analogy with more recent terminology,
let us say that a relativizable complexity class $\cald$ is leaf-definable
if $\cald$ ``admits a C-Class representation'' in the formal sense
(which we will not repeat here) 
defined by Bovet, Crescenzi, 
and 
Silvestri~(\cite{bov-cre-sil:j:uniform}, see
also~\cite{bov-cre-sil:j:sparse})
and the representation itself holds also in all relativized worlds
(under the natural extension of their work to ordered oracle access,
following exactly their discussion of relativization).  Bovet,
Crescenzi, and Silvestri~\cite{bov-cre-sil:j:uniform} prove 
that essentially all standard classes in the realm of
potentially feasible computation ``admit C-Class representations,''
they observe that these representations all relativize, and we comment
that their observation clearly holds also for ordered access.  The reason 
this is relevant is that it is easy to see that all leaf-definable
classes are sane.  Thus, the following result says that essentially
all standard complexity classes inherit every order exchange possessed
by~P.

\begin{theorem}\label{t:very-general-GeneralizedToSane}
Let ${\cal D}_1$ and ${\cal D}_2$
be sane complexity classes, and 
let ${\cal C}_1$, 
${\cal C}_2$,
${\cal C}_3$, and ${\cal C}_4$ be complexity classes. 
If ${\cal D}_1$ is robustly contained in ${\cal D}_2$ and
$\pconectwo \subseteq \p^{{\cal C}_3 : {\cal C}_4}$, then 
$${\cal D}_1^{{\cal C}_1 : {\cal C}_2} \subseteq
{\cal D}_2^{{\cal C}_3 : {\cal
C}_4}.$$
\end{theorem}

\proof
The theorem holds
via the following 
inclusion chain:
$${\cal D}_1^{{\cal C}_1 : {\cal C}_2} \stackrel{a}{\subseteq}
{\cal D}_1^{(\p^{{\cal C}_1 : {\cal C}_2})[1]} \stackrel{b}{\subseteq} 
{\cal D}_1^{(\p^{{\cal C}_3 : {\cal C}_4})[1]} \stackrel{c}{\subseteq} 
{\cal D}_2^{(\p^{{\cal C}_3 : {\cal C}_4})[1]} \stackrel{d}{\subseteq} 
{\cal D}_2^{{\cal C}_3 : 
{\cal C}_4}.$$
Inclusion (b) follows from the assumption that
$\pconectwo \subseteq \p^{{\cal C}_3 : {\cal C}_4}$ and inclusion (c)
follows from the assumption that ${\cal D}_1$ is  robustly contained in ${\cal
D}_2$. 
Inclusions (a) and (d) hold via the fact that the classes 
are sane.~\qed

\begin{corollary}\label{c:general}
Let ${\cal D}$ be any sane complexity class.
If $\pconectwo \subseteq \pctwocone$ then $${\cal D}^{{\calone} : \caltwo}
\subseteq {\cal D}^{\caltwo : \calone}.$$
\end{corollary}

We give some examples of how this can be applied.
$\bh_j$ will denote the $j$th level of the 
boolean hierarchy~\cite{cai-gun-har-hem-sew-wag-wec:j:bh1}, 
and as is standard 
DP~\cite{pap-yan:j:dp} denotes the second level of the 
boolean hierarchy.
Note that 
Bovet, 
Crescenzi, and Silvestri~\cite{bov-cre-sil:j:uniform}
have proven that BPP, UP, and PP are leaf-definable
classes.  Thus, these classes are sane. 

\begin{example} 
\begin{enumerate}
\item \label{exc}$\pp^{\np :\sigmatwo}=\pp^{\sigmatwo :\np}$.
\item \label{exa}$\bpp^{\bh_{50}:\bh_{25}} \subseteq \bpp^{\bh_{80}:\bh_{10}}=
\bpp^{\bh_{10}:\bh_{80}} \subseteq \bpp^{\bh_{25}:\bh_{38}}$.
\item \label{exb}$\up^{{\rm DP}:\bh_3}=\up^{\bh_3:{\rm DP}}=\up^{\np : \bh_3}$.
\end{enumerate}
\end{example}

Parts~\ref{exa} and~\ref{exb} of 
the example hold due to Corollary~\ref{c:general} in
light of~\cite{hem-hem-wec:jtoappear:query-order-bh}, which 
proves that 
the class of languages
computable via a polynomial-time machine given
one query to the $j$th level of the
boolean hierarchy followed by one query to the $k$th
level of the boolean hierarchy equals $\redttnp{j+2k-1}$
if $j$ is even and $k$ is odd, and equals
$\redttnp{j+2k}$ otherwise, where $\redttnp{\ell}$ equals
the class of languages that $\ell$-truth-table reduce 
to~NP sets.
Part~\ref{exc} follows, as an
application of Corollary~\ref{c:general}, from Corollary~\ref{c:sig}.

Though Theorem~\ref{t:very-general-GeneralizedToSane} 
says that {\em all\/} order exchanges of 
\p\ apply to essentially all standard 
complexity classes, it of course remains 
possible that certain path-based classes may possess additional 
order exchanges. 
For example, though Section~\ref{subs:differ} showed that \p\ ordered query 
classes create new intermediate polynomial hierarchy levels unless the 
polynomial hierarchy collapses, this clearly is not the case for \np\ 
or $\sigmak$.

\begin{theorem}
If $i \geq 1$ and $j,k \geq 0$, then
$$
{(\sigmai)}^{\sigmaj : \sigmak}=
{(\sigmai)}^{\sigmak : \sigmaj}={\Sigma_{i+\max(j,k)}^p}.$$
\end{theorem} 

\proof
Without loss of generality, suppose $j \leq k$.
Then 
${\Sigma_{i+\max(j,k)}^p}={\Sigma_{i+k}^p}={(\sigmai)}^{\sigmak}=
{(\sigmai)}^{\sigmak[1]}$ is clear in light of the quantifier
characterization of the levels of 
the polynomial hierarchy~\cite{wra:j:complete,sto:j:poly}.
Furthermore, 
${(\sigmai)}^{\sigmak[1]} \subseteq {(\sigmai)}^{\sigmaj :\sigmak} 
\subseteq {(\sigmai)}^{\sigmak} = \Sigma_{i+k}^p$  and similarly 
${(\sigmai)}^{\sigmak[1]} \subseteq {(\sigmai)}^{\sigmak :\sigmaj} 
\subseteq {(\sigmai)}^{\sigmak}= \Sigma_{i+k}^p$.~\qed

Relatedly, classes may also trivially exhibit certain equalities based on 
class-specific features. For example, it follows trivially from 
$\np \subseteq \pp$ and the (nontrivial) result of Fortnow and 
Reingold~\cite{for-rei:j:pp} regarding the $\leq_{tt}^p$ closure of \pp\ that
$\pp=\pp^{\np : \pp}=\pp^{\pp : \np}$.

Finally, as we mentioned earlier, other papers have suggested 
varying the model of this paper to include multiple queries to 
many oracles in various patterns of truth-table and ordered access.
We note that the approach of this section applies completely to 
such cases (modifying the definitions of sanity and 
leaf-definability to reflect whatever access model one is using).

{\samepage
\begin{center}
{\bf Acknowledgments}
\end{center}
\nopagebreak
\indent
We thank Gerd Wechsung for his warm encouragement,
and we thank Lance Fortnow, 
Maren Hinrichs, Leen Torenvliet, and 
Gerd Wechsung for valuable conversations and
suggestions.  
We thank anonymous referees for many helpful suggestions.
}%

\bibliographystyle{alpha}
%

\newcommand{\etalchar}[1]{$^{#1}$}

\end{document}